\documentclass[
  aps,              
  pra,              
 reprint,          
  showkeys,         
  groupedaddress,   
  ]{revtex4-2}

\setcitestyle{square,comma,sort&compress,numbers}

\usepackage[utf8]{inputenc}
\usepackage[T1]{fontenc}

\usepackage{graphicx}
\usepackage{float}

\usepackage{multirow}
\usepackage{dcolumn}

\usepackage{mathtools}
\usepackage{amssymb}
\usepackage{esint}
\usepackage{upgreek}
\usepackage{blkarray}

\usepackage[dvipsnames]{xcolor}
\usepackage{bm}
\usepackage{lmodern}
\usepackage[hidelinks]{hyperref}
\hypersetup{
  colorlinks=true,
  allcolors=blue,
  urlcolor=blue
  }
\usepackage[inline]{enumitem}
\usepackage[absolute,overlay]{textpos}

\graphicspath{{./figures/}}

\newcommand{\E}{{\rm e}}

\usepackage{xspace}
\newcommand{\ie}{\emph{i.e.}\xspace}

\newcommand{\etal}{\emph{et~al.}\xspace}

\newcommand{\cf}{\emph{cf.}\xspace}
\newcommand{\aposteriori}{\emph{a posteriori}\xspace}

\usepackage{fixme}
\fxsetup{
    status=draft,
    author=James,
    layout=inline,
    theme=color
}
\makeatletter
\renewcommand*\FXLayoutInline[3]{%
  {\@fxuseface{inline}\ignorespaces\color{blue}[#2]}}
\makeatother

\begin{document}

\title{
    \texorpdfstring{Unveiling the Burgers-Riccati physics of fast acoustic streaming}
    {Unveiling the Burgers-Riccati physics of fast acoustic streaming}
    }
    
\author{J. Orosco}
\author{J. Friend}
\altaffiliation[Corresponding author: ]{jfriend@ucsd.edu}
\affiliation{
    Medically Advanced Devices Laboratory, Center for Medical Devices\\
    Department of Mechanical and Aerospace Engineering, Jacobs School of Engineering\\
    University of California San Diego,
    La Jolla, CA 92093-0411 USA
}

\date{\today}

\begin{abstract}
    Inaccurate slow streaming models of acoustically-induced fluid flow remain in use due to the lack of a generalized alternative. The Multiscale Articulated Differentials Method (MADaM; [see co-article]) solves this problem with a complete, scale-sensitive spatiotemporal expansion. Applied to classic axial Eckart streaming, the MADaM produces, in terms of a Burgers equation, a long-sought transient solution able to accommodate frequencies and amplitudes far beyond slow streaming models. Steady streaming is governed by a corresponding Riccati equation that, when solved, produces simple expressions explaining innate features of fast acoustic streaming.
\end{abstract}

\maketitle


    \textit{Introduction.}---Flows that arise when coupling sound to fluid media---collectively referred to as ``acoustic streaming''---were first treated by Lord Rayleigh nearly one hundred fifty years ago~\cite{strutt_i._1884}. His approach is inexorably linked to order-of-magnitude separation, presumed to exist between the acoustic source's particle velocity and the much weaker net streaming velocity it generates. By ``net'' flow, we refer to fluid motions that are non-vanishing within a spatiotemporal average, $\langle\,\cdot\,\rangle_{\xi,\tau}$ taken over periodicities in acoustic space ($\xi$) and time ($\tau$) coordinates. Rayleigh's methods were later popularized by acoustofluidic pioneers including Eckart, Westervelt, and Nyborg~\cite{eckart_vortices_1948,westervelt_theory_1953,nyborg_acoustic_1965}, whom focused upon temporal separation schemes between acoustics and the induced fluid phenomena.
        
    Driven by medicine and biotechnology needs \cite{benmore_amorphization_2011,karthick_improved_2018,wu_acoustofluidic_2019,belling_acoustofluidic_2020,zhang_microliter_2021}, modern acoustofluidics research has moved far beyond classic acoustics, from the use of micro- and nano-scale fluid volumes to high-frequency (MHz and beyond) acoustic forcing~\cite{friend_microscale_2011,plaksin_intramembrane_2014,connacher_micro/nano_2018,collins_self-aligned_2018,connacher_droplet_2020}. In these systems, bulk acoustic streaming---net flow occurring in the bulk---generally attains velocities on the same order as the driving acoustics. Hence, the classic ``slow streaming'' assumption is violated and methods using it fail to properly extract the dynamics of interest. This problem has been widely recognized for at least forty years~\cite{lighthill_acoustic_1978,daru_acoustic_2017}, and yet, lacking suitable generalized alternatives, the slow streaming approach still pervades the literature~\cite{bailliet_acoustic_2001,vanneste_streaming_2011,riaud_influence_2017}.
    
    In this letter, we consider a domain with an acoustically transparent (or perfectly absorbing) distal boundary condition wherein an acoustic wave propagating to extinction produces bounded acoustic streaming. Eckart streaming, the bulk flow induced within such a system, is named after the fluid physicist who first provided a mechanistic description of its behavior~\cite{eckart_vortices_1948}. Since then, Eckart streaming has continued to receive attention due to its interesting properties and useful applications~\cite{rezk_unique_2012,tang_eckart_2017,pavlic_streaming_2021}. A theoretical framework is used here to finally remove the slow streaming constraints: the \emph{Multiscale Articulated Differentials Method} (MADaM) [see co-article]. In contrast to the solely \emph{temporal} expansions used in traditional methods, it exploits drastic \emph{spatiotemporal} scale disparities typical in microacoustofluidic systems to facilitate explicit treatment of net flow velocities commensurate with driving acoustic wave particle velocities---the \emph{fast streaming condition}. We utilize results obtained with the MADaM to comprehensively investigate fast Eckart streaming~\cite{eckart_vortices_1948}, which forms the foundation of many acoustofluidics phenomena and also routinely produces behavior known to violate the slow streaming assumption. We reveal the role of nonlinearity in precipitating monotonic spatiotemporal axial flow buildup and weakly self-similar streaming profiles. Governing transient equations are solved to explain intriguing, experimentally-observed characteristics of steady fast bulk streaming.

    \textit{Underpinnings.}---The one-dimensional model is extracted from the Navier-Stokes equations with use of nondimensional disparity parameters: $S=\omega\,t_s$, $q_p=\xi_p/x_s$, and $q_{\lambda}=(k\,x_s)^{-1}$. These are written in terms of: angular acoustic frequency, $\omega$; characteristic streaming time, $t_s$; on-source particle displacement, $\xi_p$; characteristic streaming length, $x_s$; and acoustic wavenumber, $k$. For our problem, $\mathcal{O}(S^{-1}) = \mathcal{O}(q_p) \ll \mathcal{O}(q_{\lambda})$, where the first equality is tantamount to the statement that the on-source particle velocity, $U_a=\omega\,\xi_p$, is similar in magnitude to the maximum streaming velocity, $U_s=x_s/t_s$. Or, $q_p\,S\sim1$. A complete discussion surrounding the MADaM, its application, and derivation of the one-dimensional model, all alongside its assumptions, drawbacks, and advantages, is undertaken in [see co-article].
    
    Due to the drastic disparity between the acoustic and streaming characteristic time scales---typically with $S\sim10^5$ or more---we are free to consider a regime where the acoustic field is steady and the streaming field is transient. The nondimensional steady acoustic wave depends on a complex-valued wavenumber, $\kappa = \kappa_r+\iota\,\kappa_i$, where $\iota=\sqrt{-1}$. Then $\kappa_i=(k\,\delta_a)^{-1}$ represents the amount of attenuation per unit wavelength. When $\kappa_i$ is small, the wave is approximately antisymmetric over a single period, so that spatial averages of the acoustic wave taken over an integer multiple of the wavelength are negligible. This is equivalent to $\langle\partial_{\xi}\mathcal{L}\rangle_{\xi,\tau}\approx\kappa_i\ll1$, a condition on the nondimensional acoustic Lagrangian, $\mathcal{L}$.
    
    Model equations that follow are derived on the basis of spatial (and temporal) averaging constraints, so that their validity is restricted to the range for which the acoustically-localized spatiotemporal average of the Lagrangian gradient is much less than unity. Furthermore, this condition ensures that $\kappa_i\approx\mu_l\,\omega/2\,\rho_0\,c^2$, where $\mu_l=\mu_s(4/3+\mu_v/\mu_s)$ is the longitudinal viscosity ($\mu_s$ is the shear viscosity and $\mu_v$ is the volume viscosity), $\rho_0$ is the fluid density, and $c$ is the fluid sound speed. The fast streaming equations investigated in this letter remain valid for acoustic frequencies ranging from at least 10\,kHz to 25\,GHz based on continuum and spatial averaging limitations.

    \textit{Transient Burgers flow.}---In the presence of a perfectly acoustically absorbing distal boundary, an acoustic wave generated at the origin will propagate continuously, unhindered, while streaming flow must adhere to a no-slip condition. Essential physics underlying fast axial flow in this Eckart configuration is succinctly captured with the forced viscous Burgers equation
        %
        \begin{align}\label{eq:transient_burgers_pde}
            \partial_tu+u\,\partial_xu = \mu\,\partial_x^2\,u+\eta_m^{-1}f_R(x),
        \end{align}
    where $\mu=q_{\lambda}/R$ is the nondimensional viscosity, $R=\rho_0\,x_s\,U_s/\mu_l$ is a Reynolds number for the streaming flow, and $f_R(x)=-\langle u^{(a)}\partial_xu^{(a)}\rangle_{\xi,\tau}$ is the Reynolds stress forcing, obtained as an acoustically-localized spatiotemporal average of the advected acoustic wave. We define the extent of acoustic energy transduction, $\eta(x)=(u(x)/U_a)^2$, and the maximum streaming conversion efficiency, $\eta_m=\max_x\eta(x)=(q_p\,S)^{-2}=U_s^2/U_a^2$, in terms of the ratio of the maximum streaming velocity, $U_s$, to the source's particle velocity, $U_a$. Its appearance in Eq.\,\eqref{eq:transient_burgers_pde} underscores the role of Reynolds stress as a transduction mechanism between the two fields.
    
\begin{figure*}[htb!]
            \begin{center}
                \includegraphics[height=9.3cm]{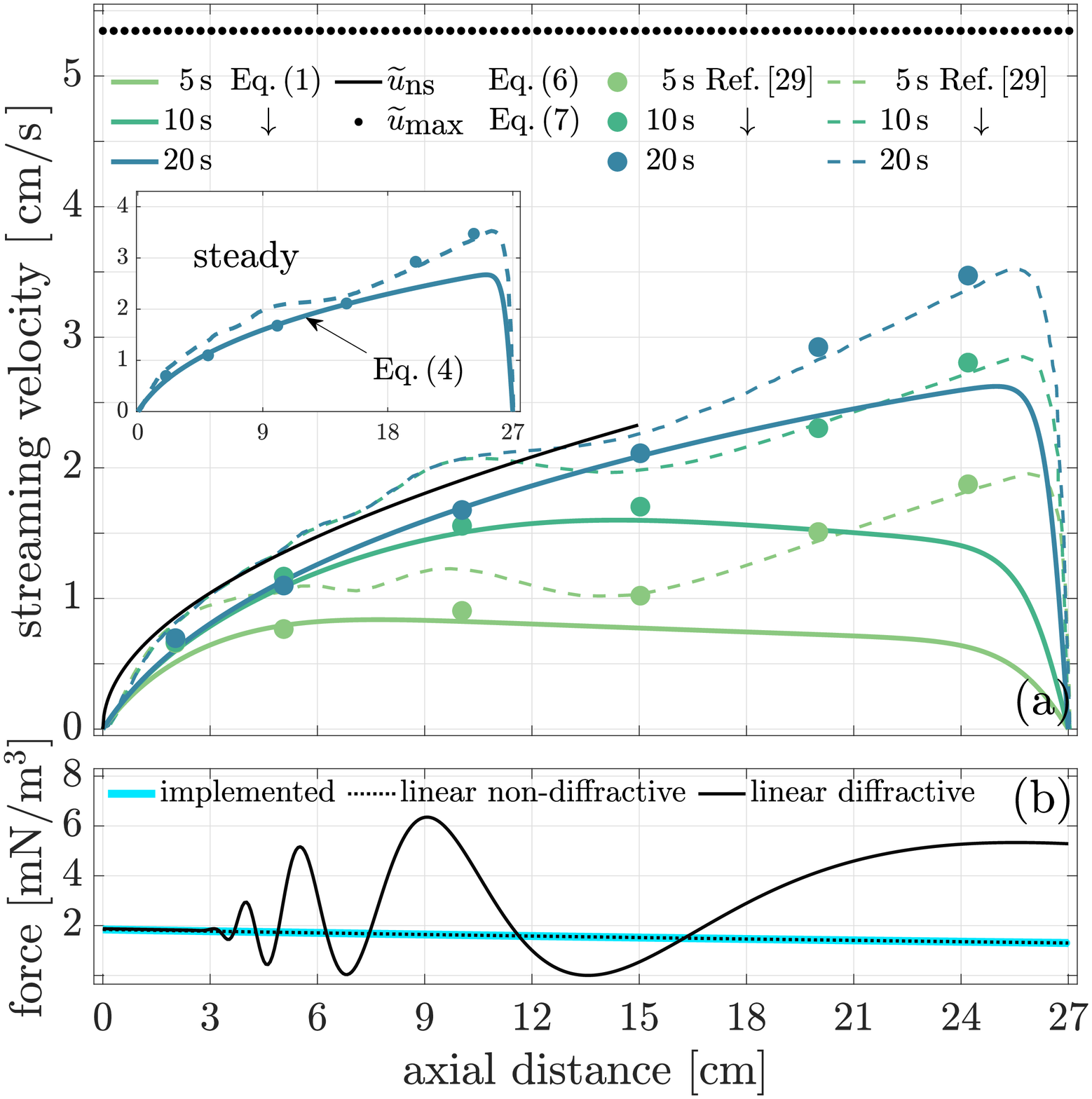}
                \includegraphics[height=9.3cm]{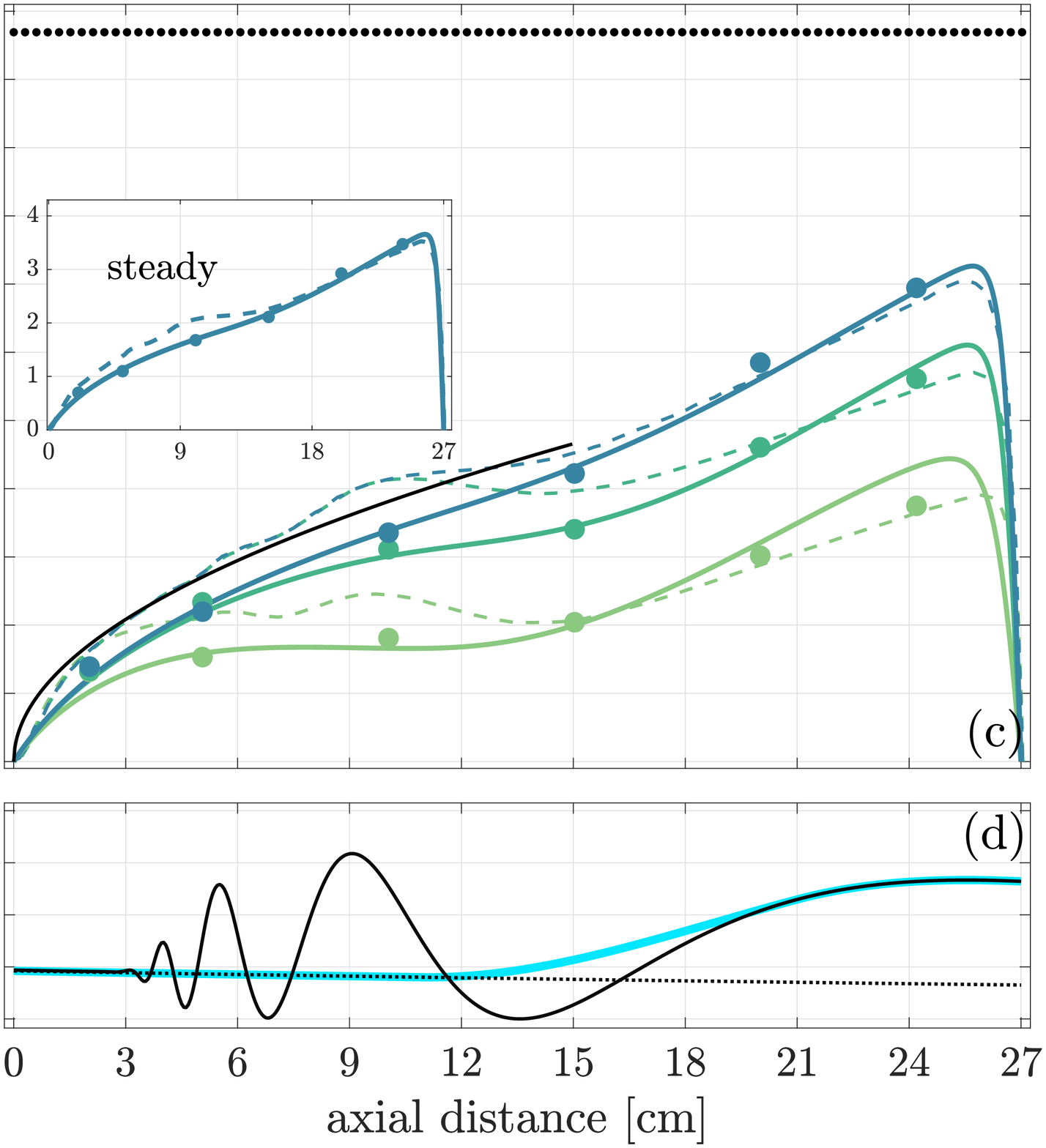}
                \caption{Large-amplitude axial Eckart streaming. (a) Viscous Burgers streaming model with $\mu\approx0.018$ and (b) Reynolds stress derived from a linear, non-diffractive pressure field. (c) Viscous Burgers streaming model with $\mu\approx0.013$ and (d) composite Reynolds stress derived from a linear non-diffractive pressure field in the leading half of the domain and a linear diffractive pressure field in the latter half of the domain. All other parameters follow from Ref.\,\cite{kamakura_time_1996}. Steady profile evolution---which starts at the source and proceeds monotonically toward the distal boundary---leads to spatially progressive profile dependence. Thus, forcing in the leading section of the model of Ref.\,\cite{kamakura_time_1996} cannot be altered without significantly perturbing model fidelity in the latter domain half.}
                \label{fig:f01_2d_kama_transient}
            \end{center}
        \end{figure*}

    Broader interpretation of Eq.\,\eqref{eq:transient_burgers_pde} is possible by defining the streaming potential, $h$, such that $u=-\partial_xh$. Accordingly, the Burgers streaming equation may be rewritten as
        %
        \begin{align}\label{eq:kardar_parisi_zhang}
            \partial_th=\nu\,\partial_x^2h+\frac{\Lambda}{2}(\partial_xh)^2+F.
        \end{align}
    When $F$ is a zero-mean Gaussian white noise, Eq.\,\eqref{eq:kardar_parisi_zhang} is known as the Kardar-Parisi-Zhang (KPZ) equation~\cite{kardar_dynamic_1986}. The KPZ equation is a stochastic growth model for the interfacial height of a body under random deposition. It has been used for describing tumor growth and ballistic deposition, among a variety of other processes~\cite{meakin_ballistic_1986,sasamoto_one-dimensional_2010,santalla_eden_2018,nahum_quantum_2017}. Points of nondifferentiability in randomly ``roughened'' height profiles produce sawtooth gradients characteristic of Burgers shock front modeling~\cite{burgers_nonlinear_1974}. In the present setting, with deterministic forcing $\partial_xF=f_R$, Eq.\,\eqref{eq:kardar_parisi_zhang} describes an axial buildup potential, the gradient of which returns the streaming velocity. In absence of forcing and with arbitrary initial profile, Eq.\,\eqref{eq:kardar_parisi_zhang} generates solutions that asymptotically evolve toward smooth steady profiles under layered growth~\cite{kardar_dynamic_1986}. When subjected to forcing via Reynolds stress, a combination of these characteristics is observed in formation of the ``shark fin'' (\ie, rounded sawtooth) profile typifying axial Eckart streaming (Fig.\,\ref{fig:f01_2d_kama_transient}). Layered evolution of this profile is directly analogous to the noted deposition processes.

\begin{figure*}[htb!]
            \begin{center}
                \includegraphics[height=8.1 cm]{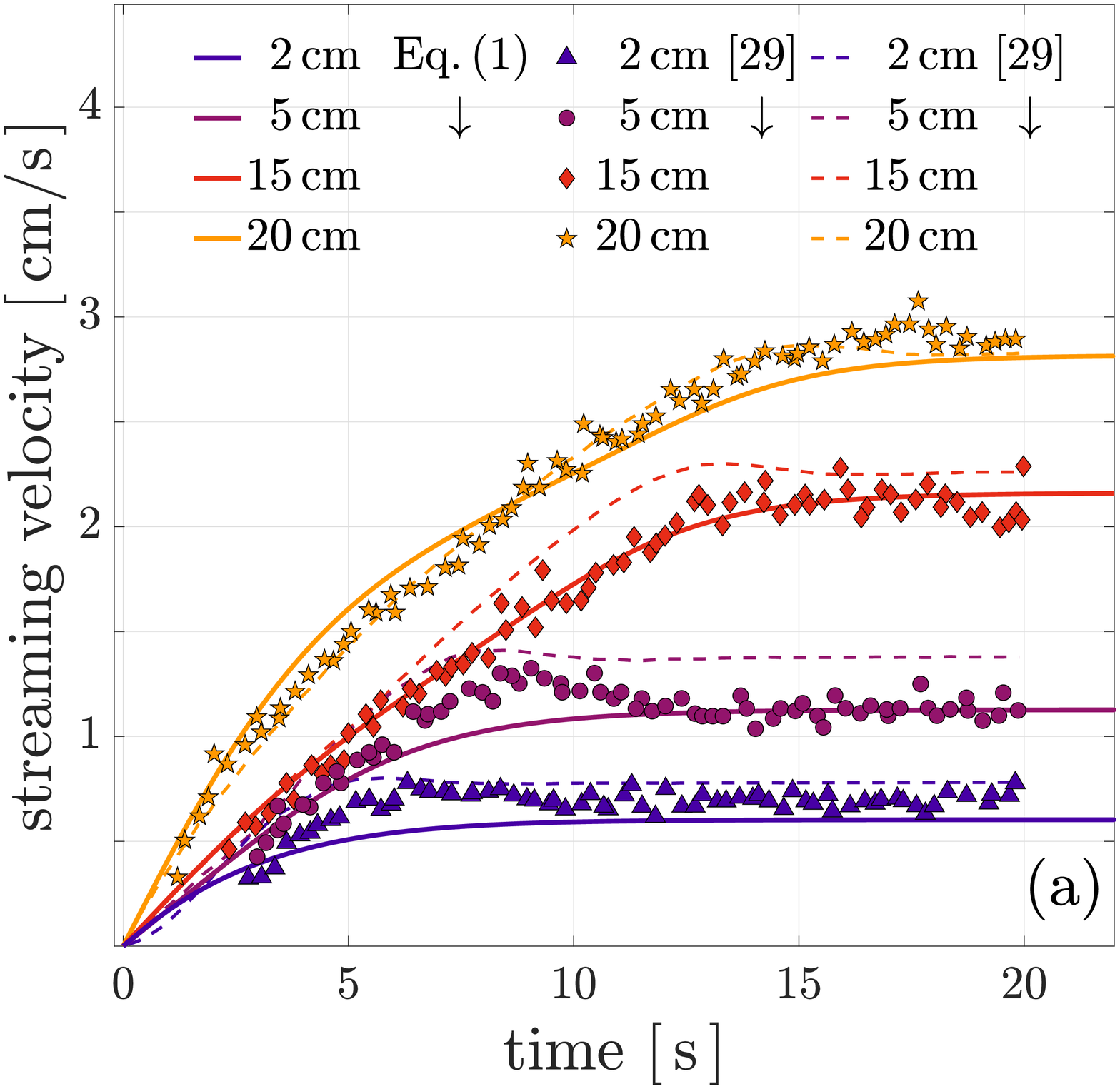}
                \includegraphics[height=8.1 cm]{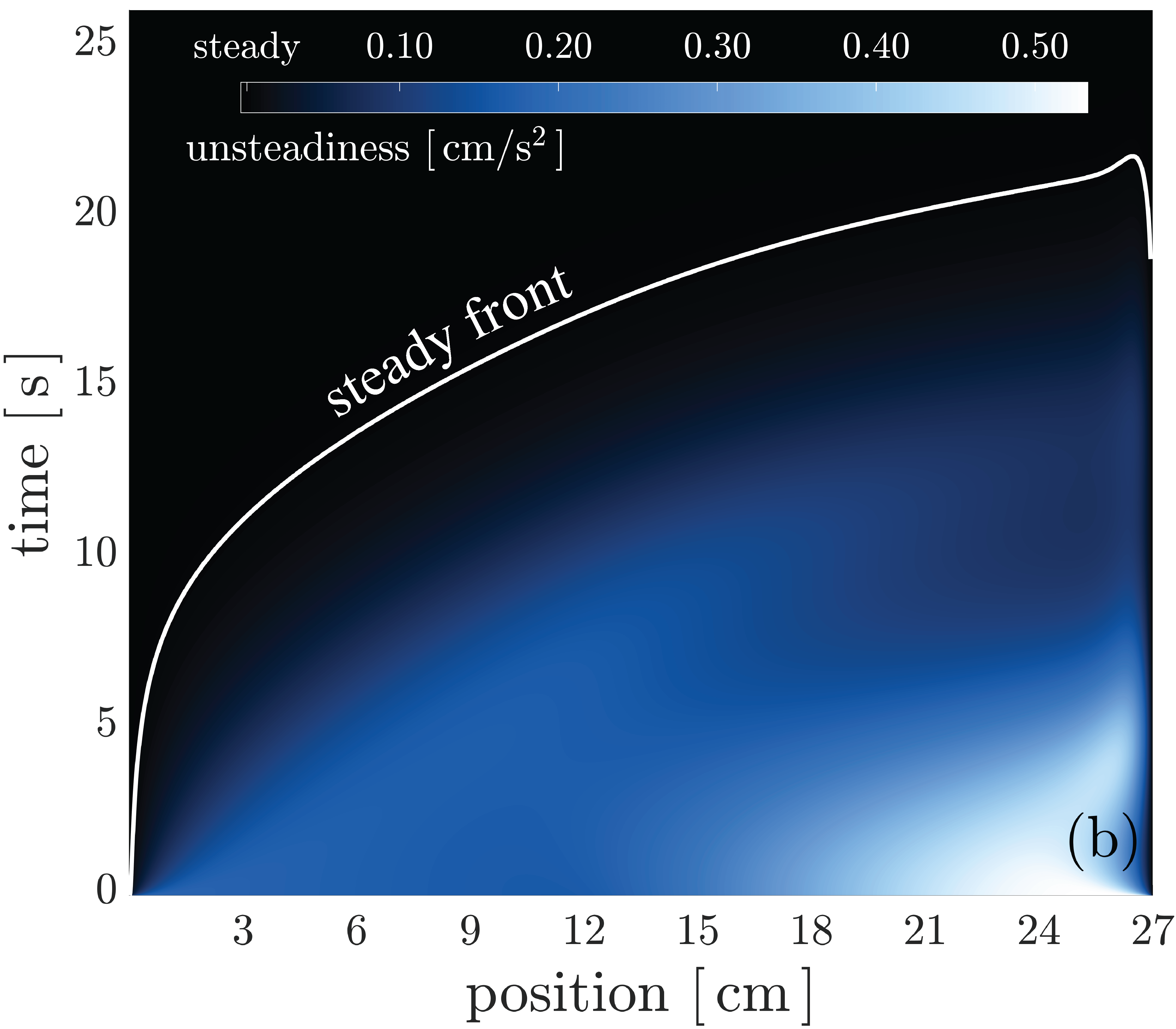}
                \caption{Large-amplitude Eckart streaming from onset to steady state corresponding to the model in Fig.\,\href{fig:f01_2d_kama_transient}{1(c),(d)}. (a) The Burgers streaming model accurately describes fundamental characteristics of transient growth of the phenomena from quiescent conditions. (b) A spatiotemporal map of flow unsteadiness reveals a steady front (white solid line) that develops monotonically due to uniform streaming flow layering. The steady front is defined as $1\,$\% of maximum flow unsteadiness, which is  $\widetilde{\partial}_t\widetilde{u}\approx0.005\,$cm/s$^2$ here.}
                \label{fig:f02_kama_trans_decay}
            \end{center}
        \end{figure*}

    Direct comparison of Eq.\,\eqref{eq:transient_burgers_pde} with Eckart streaming-driven transient and steady characterizations in Kamakura \etal~\cite{kamakura_time_1996} is made in Figs.\,\ref{fig:f01_2d_kama_transient} and \ref{fig:f02_kama_trans_decay}. For solving the transient PDE Eq.\,\eqref{eq:transient_burgers_pde}, we employ the finite element method with the \emph{FEniCS Project} suite of components~\cite{alnaes_unified_2009,alnaes_fenics_2015,alnaes_unified_2014,kirby_algorithm_2004,kirby_compiler_2006,logg_automated_2012,logg_dolfin_2010,olgaard_optimizations_2010}. Kamakura's modeling results were obtained by numerical integration of the axisymmetric incompressible Navier-Stokes equations under a standard field partition (\cf\ Ref.\,\cite{rudenko_theoretical_1977}, p.\,192). Kamakura's acoustic forcing is derived from sound pressure, where the latter is modeled with the Khokhlov-Zabolotskaya-Kuznetsov (KZK) equation~\cite{zabolotskaya_quasiplane_1969,kuznetsov_equations_1971,novikov_nonlinear_1987}. In this study, we obtain the continuous wave solution to the KZK equation with the \emph{Fast Object-Oriented C++ Ultrasound Simulator (FOCUS) \textsc{Matlab} Toolbox}~\cite{mcgough_efficient_2004,chen_2d_2008,mcgough_fast_2021}.
    
    Figure\,\ref{fig:f01_2d_kama_transient} subplots depict two scenarios. Kamakura's model (observable in either plot) employs Reynolds stress derived from a nonlinear, diffractive KZK pressure field. The model amplitude and shape poorly represent the streaming profile over the leading half of the domain. For circular plane transducers, the domain of validity of the KZK equation is inherited from the quasi-optical approximation upon which it is based. This begins at a distance $d_v=a\,(k\,a)^{1/3}/2\approx2.6\,$cm from the source (\cf\ Ref.\,\cite{novikov_nonlinear_1987}, p.\,50), where $a$ is the transducer aperture diameter. The spurious region in Kamakura's model extends well into the KZK equation domain of validity.
    
    Results in Fig.\,\href{fig:f01_2d_kama_transient}{1(a)} are obtained with Eq.\,\eqref{eq:transient_burgers_pde} assuming a linear, non-diffractive plane wave, inducing the Reynolds stress $f_R(x)=(\overline{\alpha}/2)\,\E^{-2\,\overline{\alpha}\,x}$, where the latter is plotted in Fig.\,\href{fig:f01_2d_kama_transient}{1(b)}. The nondimensional attenuation coefficient is $\overline{\alpha}=\kappa_i/q_{\lambda}$. Agreement of the model with Kamakura's observations over the leading half of the domain is remarkable. In the latter domain half, however, the model diverges from observations. In Fig.\,\href{fig:f01_2d_kama_transient}{1(c)}, we show Eq.\,\eqref{eq:transient_burgers_pde} driven by a Reynolds stress composed of the linear, non-diffractive solution in the near field and of the linear, diffractive solution---that is, the solution to the KZK equation with nonlinearity parameter set to zero---in the far field. The two forcing regimes are joined by a spline with second-order smoothness, as shown in Fig.\,\href{fig:f01_2d_kama_transient}{1(d)}, before being applied to the transient Burgers streaming model. The root mean square relative error of Kamakura's model is $23\,\%$, $18\,\%$, and $15\,\%$, at $5\,$s, $10\,$s, and $20\,$s, respectively. For the concise Burgers model with modified Reynolds stress, the values are $12\,\%$, $5\,\%$, and $6\,\%$, respectively.
    
    Kamakura's results seem to imply that by adjusting his model's near-source forcing conditions, better overall agreement with the observations can be achieved. This assessment is inaccurate: upstream forcing significantly impacts downstream flow, while forcing near the distal boundary negligibly affects streaming near the source. Streaming flow steadiness develops first near the source and progresses monotonically toward the distal boundary (see Fig.\,\href{fig:f02_kama_trans_decay}{2(b)} and associated discussion). The modified Reynolds stress better aligns with the model physics, since near-source, non-diffractive forcing is left intact while forcing over the remainder of the domain is brought into agreement with experimentally observed diffractive enhancement. Hence, progressive spatial dependence as a function of monotonic flow development provides strong \aposteriori support for veracity of the concise Burgers model.
        
    Streaming field transience is more thoroughly investigated in Fig.\,\ref{fig:f02_kama_trans_decay}. In Fig.\,\href{fig:f02_kama_trans_decay}{2(a)}, diffractive enhancement causes early departure of distal flow buildup, and this enhancement persists to steady state. Near the source, however, diffractive enhancement is transient and rapidly decays toward the steady equivalent of a non-diffractively forced flow. The steady flow profile develops in a layered manner that is to date unexplored to our knowledge. The origin of this phenomenon is revealed by the relationship between Eqs.\,\eqref{eq:transient_burgers_pde} and \eqref{eq:kardar_parisi_zhang}. At each point along the profile, the flow velocity locally increases over time to a steady value after a unique amount of elapsed time. Counterintuitively, steady flow is first established near the dynamic boundary. Over time, flow steadiness exponentially advances along the domain length before transitioning to a nearly linear growth rate over the latter third of the domain. Steady flow is last achieved close to the distal boundary. Spatiotemporal monotonic steady state development is directly evident in the unsteadiness map of Fig.\,\href{fig:f02_kama_trans_decay}{2(b)}, where we define a steady front at $1\,$\% of maximum flow unsteadiness, $\widetilde{\partial}_t\widetilde{u}\approx0.005\,$cm/s$^2$ here.

    \textit{Steady Riccati flow.}---When analysing acoustofluidic systems, one's objective is often recovery of an equation for \emph{steady} streaming.  Setting the unsteady term in Eq.\,\eqref{eq:transient_burgers_pde} to zero, integrating over the domain, and rearranging the result, we obtain Riccati's equation~\cite{riccati_animadversiones_1724} for steady fast Eckart streaming
        %
        \begin{align}\label{eq:steady_riccati_ode}
            &q_{\lambda}\,(\partial_x\,u-\partial_x\,u|_{x=0})-\frac{R}{2}\,u^2 \notag \\
            &\hspace{1.5cm}= \frac{R}{\eta_m}\int^x_0\langle u^{(a)}\partial_xu^{(a)}\rangle_{\xi,\tau}\,dx.
        \end{align}
    The role of $q_{\lambda}=(k\,x_s)^{-1}$ is revealed here. With $q_{\lambda}\sim1$, all flow features are selected and the full equation is expressed. However, with $q_{\lambda}\ll1$, only those flow features with dimension $x_s$ much larger than the wavelength will be retained. Since the viscous term is directly related to the boundary layer, and the boundary layer thickness, $\delta_v=\sqrt{2\,\mu_l/\rho_0\,\omega}$, is much smaller than the wavelength (\ie, $\lambda/\delta_v\gg1$), the latter condition on $q_{\lambda}$ tends to exclude the parenthetical term in Eq.\,\eqref{eq:steady_riccati_ode}. This corresponds to loss of the distal boundary condition---that is, the boundary layer is discarded---so that the resulting algebraic expression can be solved to approximate bulk flow. Evidently, then, bulk axial flow in such a scenario is not dependent on the streaming Reynolds number. Indeed, this property of bulk Eckart streaming leads to weak self-similarity, as made evident in the following inviscid analysis and as observed experimentally~\cite{moudjed_near-field_2015}.
    
   When $u^{(a)}$ is a linear, non-diffractive acoustic wave, a solution method exists for the steady viscous problem that involves transforming Eq.\,\eqref{eq:steady_riccati_ode} into a second order linear equation~\cite{ince_ordinary_1956}. The result is
        %
        \begin{subequations}\label{eq:viscous_riccati_solution}
            \begin{align}
                u_{\mbox{\tiny visc}}&=\frac{\partial_x\,\phi}{c_s\,\phi},\\
                \phi&=I_{\beta}(h)+c_{\phi}I_{-\beta}(h),\\
                c_{\phi}&=-\frac{I_{\beta+1}(h_0)+I_{\beta-1}(h_0)}{I_{-(\beta+1)}(h_0)+I_{-(\beta-1)}(h_0)},\\
                h&=h_0\,\text{e}^{-\overline{\alpha}\,x},
            \end{align}
        \end{subequations}
    where $c_s = -(2\,\mu)^{-1}$, $c_f=-c_s/2\,\eta$, $\beta=h_0\sqrt{(\partial_x\,u_{\mbox{\tiny visc}}|_{x=0}-c_f)/c_f}$, and $h_0=\sqrt{(c_s\,c_f)/\overline{\alpha}^2}$. Here $I$ denotes the modified Bessel's function of the first kind. The steady source flow gradient, $\partial_x\,u_{\mbox{\tiny visc}}|_{x=0}$, has a unique value corresponding to satisfaction of the homogeneous distal boundary condition.
    
    Equation\,\eqref{eq:viscous_riccati_solution}, plotted in the  Fig.\,\href{fig:f01_2d_kama_transient}{1(a)} inset, overlaps a corresponding solution to the transient Burgers equation as $t\uparrow\infty$. When evaluating  Eq.\,\eqref{eq:viscous_riccati_solution}---where arbitrary precision operation is desirable for computing linear combinations of Bessel functions---we have employed the \emph{Advanpix Multiprecision Computing \textsc{Matlab} Toolbox}~\cite{holoborodko_multiprecision_2020}. In general, Eq.\,\eqref{eq:viscous_riccati_solution} generates a shark fin profile where flow in the fluid bulk is determined primarily by inertia (and hence, nonlinearity) and partially mediated by viscosity (Fig.\,\ref{fig:f01_2d_kama_transient}).
    
    We obtain the steady, bulk solution with $\mu\approx0$ (\ie, $q_{\lambda}\downarrow0$ or $R\uparrow\infty$), so that by inspection of Eq.\,\eqref{eq:transient_burgers_pde} \mbox{$u_{\mbox{\tiny bulk}} = [\tfrac{1}{2\,\eta}(1-\exp(-2\,\overline{\alpha}\,x))]^{1/2}$}. Its dimensional form is written
        %
        \begin{align}\label{eq:inviscid_streaming}
            \widetilde{u}_{\tiny \mbox{bulk}}=U_a\sqrt{\tfrac{1}{2}\sum_{n=1}^{\infty}\tfrac{(2\,\alpha\,\widetilde{x})^n}{n!}},
        \end{align}
    where \mbox{$\alpha=\kappa_i\,k$} is the ``true'' absorption coefficient. If $\widetilde{x}_\text{ns}$ is a point sufficiently close to the source, then $2\,\alpha\,\widetilde{x}_\text{ns}\ll1$ and
        %
        \begin{align}\label{eq:near_source_approximation}
            \widetilde{u}_\text{ns} = U_a\sqrt{\alpha\,\widetilde{x}_\text{ns}},
        \end{align}
    describes near-source inviscid streaming. This is plotted in Fig.\,\ref{fig:f01_2d_kama_transient}, where one observes that for Kamakura's observations, the flow profile follows an approximate square root dependence over the leading half of the domain. We deduce from Eq.\,\eqref{eq:near_source_approximation} that, near the source, acoustic wave to streaming flow transduction is linearly dependent on axial distance by a factor equivalent to the attenuation coefficient, $\eta(\widetilde{x}_\text{ns})\approx\alpha\,\widetilde{x}_\text{ns}$.
    
    From Eq.\,\eqref{eq:inviscid_streaming}, one determines the maximum achievable streaming velocity:
        %
        \begin{align}\label{eq:bulk_streaming_speed_limit}
            \max_{\forall}\widetilde{u}=U_a/\sqrt{2},
        \end{align}
    which can evidently be realized in the limiting case of an infinite attenuation coefficient---which itself is a quadratic function of forcing frequency---or very far from the source in the limiting case of an infinite flow domain. The relevant value is indicated in Fig.\,\ref{fig:f01_2d_kama_transient}. The design implication of the foregoing results is that the only means of achieving maximal velocity output under domain length restriction is by increasing frequency. This is emphasized by the fact that Eq.\,\eqref{eq:inviscid_streaming} attains its maximum as a Gaussian function of increasing frequency. Interested readers are directed to [see co-article], where a broad survey of the applicability of Eq.\,\eqref{eq:bulk_streaming_speed_limit} is undertaken, and effects of forcing frequency on steady and transient flow characteristics are explored in greater detail.
        
    Dividing Eq.\,\eqref{eq:bulk_streaming_speed_limit} through by $U_a$, squaring both sides, and substituting the streaming conversion efficiency, we arrive at
        %
        \begin{align}
            \max_{\forall}\left(\frac{|\widetilde{u}^{(s)}|}{U_a}\right)^2=\max_{\forall}\eta_m=\frac{1}{2},
        \end{align}
    so that the streaming law expressed by Eq.\,\eqref{eq:bulk_streaming_speed_limit} may apparently be interpreted as a fundamental limit on acoustic energy transduction efficiency of $50\,\%$. Here our notation implies a maximum ``for all'' possible system configurations described by this model. This limit is \emph{entirely independent of constitutive parameters}.

\begin{acknowledgments}
    The work presented here was generously supported by a SERF research grant to J. Friend from the W. M. Keck Foundation. He is furthermore grateful for the support of this work by the Office of Naval Research (Grant No. 12368098). J. Orosco is thankful for support provided by the University of California's Presidential Postdoctoral Fellowship Program.
\end{acknowledgments}

\bibliographystyle{apsrev4-2} 
\bibliography{1d_burgers}

\begin{thebibliography}{48}%
\makeatletter
\providecommand \@ifxundefined [1]{%
 \@ifx{#1\undefined}
}%
\providecommand \@ifnum [1]{%
 \ifnum #1\expandafter \@firstoftwo
 \else \expandafter \@secondoftwo
 \fi
}%
\providecommand \@ifx [1]{%
 \ifx #1\expandafter \@firstoftwo
 \else \expandafter \@secondoftwo
 \fi
}%
\providecommand \natexlab [1]{#1}%
\providecommand \enquote  [1]{``#1''}%
\providecommand \bibnamefont  [1]{#1}%
\providecommand \bibfnamefont [1]{#1}%
\providecommand \citenamefont [1]{#1}%
\providecommand \href@noop [0]{\@secondoftwo}%
\providecommand \href [0]{\begingroup \@sanitize@url \@href}%
\providecommand \@href[1]{\@@startlink{#1}\@@href}%
\providecommand \@@href[1]{\endgroup#1\@@endlink}%
\providecommand \@sanitize@url [0]{\catcode `\\12\catcode `\$12\catcode
  `\&12\catcode `\#12\catcode `\^12\catcode `\_12\catcode `\%12\relax}%
\providecommand \@@startlink[1]{}%
\providecommand \@@endlink[0]{}%
\providecommand \url  [0]{\begingroup\@sanitize@url \@url }%
\providecommand \@url [1]{\endgroup\@href {#1}{\urlprefix }}%
\providecommand \urlprefix  [0]{URL }%
\providecommand \Eprint [0]{\href }%
\providecommand \doibase [0]{https://doi.org/}%
\providecommand \selectlanguage [0]{\@gobble}%
\providecommand \bibinfo  [0]{\@secondoftwo}%
\providecommand \bibfield  [0]{\@secondoftwo}%
\providecommand \translation [1]{[#1]}%
\providecommand \BibitemOpen [0]{}%
\providecommand \bibitemStop [0]{}%
\providecommand \bibitemNoStop [0]{.\EOS\space}%
\providecommand \EOS [0]{\spacefactor3000\relax}%
\providecommand \BibitemShut  [1]{\csname bibitem#1\endcsname}%
\let\auto@bib@innerbib\@empty
\bibitem [{\citenamefont {Strutt}(1884)}]{strutt_i._1884}%
  \BibitemOpen
  \bibfield  {author} {\bibinfo {author} {\bibfnamefont {J.~W.}\ \bibnamefont
  {Strutt}},\ }\href {https://doi.org/10.1098/rstl.1884.0002} {\bibfield
  {journal} {\bibinfo  {journal} {Philos. T. Roy. Soc. Lon.}\ }\textbf
  {\bibinfo {volume} {175}},\ \bibinfo {pages} {1} (\bibinfo {year}
  {1884})}\BibitemShut {NoStop}%
\bibitem [{\citenamefont {Eckart}(1948)}]{eckart_vortices_1948}%
  \BibitemOpen
  \bibfield  {author} {\bibinfo {author} {\bibfnamefont {C.}~\bibnamefont
  {Eckart}},\ }\href {https://doi.org/10.1103/PhysRev.73.68} {\bibfield
  {journal} {\bibinfo  {journal} {Phys. Rev.}\ }\textbf {\bibinfo {volume}
  {73}},\ \bibinfo {pages} {68} (\bibinfo {year} {1948})}\BibitemShut {NoStop}%
\bibitem [{\citenamefont {Westervelt}(1953)}]{westervelt_theory_1953}%
  \BibitemOpen
  \bibfield  {author} {\bibinfo {author} {\bibfnamefont {P.~J.}\ \bibnamefont
  {Westervelt}},\ }\href {https://doi.org/10.1121/1.1907009} {\bibfield
  {journal} {\bibinfo  {journal} {J. Acoust. Soc. Am.}\ }\textbf {\bibinfo
  {volume} {25}},\ \bibinfo {pages} {60} (\bibinfo {year} {1953})}\BibitemShut
  {NoStop}%
\bibitem [{\citenamefont {Nyborg}(1965)}]{nyborg_acoustic_1965}%
  \BibitemOpen
  \bibfield  {author} {\bibinfo {author} {\bibfnamefont {W.~L.~M.}\
  \bibnamefont {Nyborg}},\ }in\ \href
  {https://doi.org/10.1016/B978-0-12-395662-0.50015-1} {\emph {\bibinfo
  {booktitle} {Physical {{Acoustics}}}}},\ \bibinfo {series} {Properties of
  {{Polymers}} and {{Nonlinear Acoustics}}}, Vol.~\bibinfo {volume} {2},\
  \bibinfo {editor} {edited by\ \bibinfo {editor} {\bibfnamefont {W.~P.}\
  \bibnamefont {Mason}}}\ (\bibinfo  {publisher} {{Academic Press}},\ \bibinfo
  {address} {{New York}},\ \bibinfo {year} {1965})\ pp.\ \bibinfo {pages}
  {265--331}\BibitemShut {NoStop}%
\bibitem [{\citenamefont {Benmore}\ and\ \citenamefont
  {Weber}(2011)}]{benmore_amorphization_2011}%
  \BibitemOpen
  \bibfield  {author} {\bibinfo {author} {\bibfnamefont {C.~J.}\ \bibnamefont
  {Benmore}}\ and\ \bibinfo {author} {\bibfnamefont {J.~K.~R.}\ \bibnamefont
  {Weber}},\ }\href {https://doi.org/10.1103/PhysRevX.1.011004} {\bibfield
  {journal} {\bibinfo  {journal} {Phys. Rev. X}\ }\textbf {\bibinfo {volume}
  {1}},\ \bibinfo {pages} {011004} (\bibinfo {year} {2011})}\BibitemShut
  {NoStop}%
\bibitem [{\citenamefont {Karthick}\ and\ \citenamefont
  {Sen}(2018)}]{karthick_improved_2018}%
  \BibitemOpen
  \bibfield  {author} {\bibinfo {author} {\bibfnamefont {S.}~\bibnamefont
  {Karthick}}\ and\ \bibinfo {author} {\bibfnamefont {A.~K.}\ \bibnamefont
  {Sen}},\ }\href {https://doi.org/10.1103/PhysRevApplied.10.034037} {\bibfield
   {journal} {\bibinfo  {journal} {Phys. Rev. Applied}\ }\textbf {\bibinfo
  {volume} {10}},\ \bibinfo {pages} {034037} (\bibinfo {year}
  {2018})}\BibitemShut {NoStop}%
\bibitem [{\citenamefont {Wu}\ \emph {et~al.}(2019)\citenamefont {Wu},
  \citenamefont {Ozcelik}, \citenamefont {Rufo}, \citenamefont {Wang},
  \citenamefont {Fang},\ and\ \citenamefont
  {Jun~Huang}}]{wu_acoustofluidic_2019}%
  \BibitemOpen
  \bibfield  {author} {\bibinfo {author} {\bibfnamefont {M.}~\bibnamefont
  {Wu}}, \bibinfo {author} {\bibfnamefont {A.}~\bibnamefont {Ozcelik}},
  \bibinfo {author} {\bibfnamefont {J.}~\bibnamefont {Rufo}}, \bibinfo {author}
  {\bibfnamefont {Z.}~\bibnamefont {Wang}}, \bibinfo {author} {\bibfnamefont
  {R.}~\bibnamefont {Fang}},\ and\ \bibinfo {author} {\bibfnamefont
  {T.}~\bibnamefont {Jun~Huang}},\ }\href
  {https://doi.org/10.1038/s41378-019-0064-3} {\bibfield  {journal} {\bibinfo
  {journal} {Microsyst. Nanoeng.}\ }\textbf {\bibinfo {volume} {5}},\ \bibinfo
  {pages} {1} (\bibinfo {year} {2019})}\BibitemShut {NoStop}%
\bibitem [{\citenamefont {Belling}\ \emph {et~al.}(2020)\citenamefont
  {Belling}, \citenamefont {Heidenreich}, \citenamefont {Tian}, \citenamefont
  {Mendoza}, \citenamefont {Chiou}, \citenamefont {Gong}, \citenamefont {Chen},
  \citenamefont {Young}, \citenamefont {Wattanatorn}, \citenamefont {Park},
  \citenamefont {Scarabelli}, \citenamefont {Chiang}, \citenamefont
  {Takahashi}, \citenamefont {Young}, \citenamefont {Stieg}, \citenamefont
  {Oliveira}, \citenamefont {Huang}, \citenamefont {Weiss},\ and\ \citenamefont
  {Jonas}}]{belling_acoustofluidic_2020}%
  \BibitemOpen
  \bibfield  {author} {\bibinfo {author} {\bibfnamefont {J.~N.}\ \bibnamefont
  {Belling}}, \bibinfo {author} {\bibfnamefont {L.~K.}\ \bibnamefont
  {Heidenreich}}, \bibinfo {author} {\bibfnamefont {Z.}~\bibnamefont {Tian}},
  \bibinfo {author} {\bibfnamefont {A.~M.}\ \bibnamefont {Mendoza}}, \bibinfo
  {author} {\bibfnamefont {T.-T.}\ \bibnamefont {Chiou}}, \bibinfo {author}
  {\bibfnamefont {Y.}~\bibnamefont {Gong}}, \bibinfo {author} {\bibfnamefont
  {N.~Y.}\ \bibnamefont {Chen}}, \bibinfo {author} {\bibfnamefont {T.~D.}\
  \bibnamefont {Young}}, \bibinfo {author} {\bibfnamefont {N.}~\bibnamefont
  {Wattanatorn}}, \bibinfo {author} {\bibfnamefont {J.~H.}\ \bibnamefont
  {Park}}, \bibinfo {author} {\bibfnamefont {L.}~\bibnamefont {Scarabelli}},
  \bibinfo {author} {\bibfnamefont {N.}~\bibnamefont {Chiang}}, \bibinfo
  {author} {\bibfnamefont {J.}~\bibnamefont {Takahashi}}, \bibinfo {author}
  {\bibfnamefont {S.~G.}\ \bibnamefont {Young}}, \bibinfo {author}
  {\bibfnamefont {A.~Z.}\ \bibnamefont {Stieg}}, \bibinfo {author}
  {\bibfnamefont {S.~D.}\ \bibnamefont {Oliveira}}, \bibinfo {author}
  {\bibfnamefont {T.~J.}\ \bibnamefont {Huang}}, \bibinfo {author}
  {\bibfnamefont {P.~S.}\ \bibnamefont {Weiss}},\ and\ \bibinfo {author}
  {\bibfnamefont {S.~J.}\ \bibnamefont {Jonas}},\ }\href
  {https://doi.org/10.1073/pnas.1917125117} {\bibfield  {journal} {\bibinfo
  {journal} {PNAS}\ }\textbf {\bibinfo {volume} {117}},\ \bibinfo {pages}
  {10976} (\bibinfo {year} {2020})}\BibitemShut {NoStop}%
\bibitem [{\citenamefont {Zhang}\ \emph {et~al.}(2021)\citenamefont {Zhang},
  \citenamefont {{Zuniga-Hertz}}, \citenamefont {Zhang}, \citenamefont
  {Gopesh}, \citenamefont {Fannon}, \citenamefont {Wang}, \citenamefont {Wen},
  \citenamefont {Patel},\ and\ \citenamefont {Friend}}]{zhang_microliter_2021}%
  \BibitemOpen
  \bibfield  {author} {\bibinfo {author} {\bibfnamefont {N.}~\bibnamefont
  {Zhang}}, \bibinfo {author} {\bibfnamefont {J.~P.}\ \bibnamefont
  {{Zuniga-Hertz}}}, \bibinfo {author} {\bibfnamefont {E.~Y.}\ \bibnamefont
  {Zhang}}, \bibinfo {author} {\bibfnamefont {T.}~\bibnamefont {Gopesh}},
  \bibinfo {author} {\bibfnamefont {M.~J.}\ \bibnamefont {Fannon}}, \bibinfo
  {author} {\bibfnamefont {J.}~\bibnamefont {Wang}}, \bibinfo {author}
  {\bibfnamefont {Y.}~\bibnamefont {Wen}}, \bibinfo {author} {\bibfnamefont
  {H.~H.}\ \bibnamefont {Patel}},\ and\ \bibinfo {author} {\bibfnamefont
  {J.}~\bibnamefont {Friend}},\ }\href {https://doi.org/10.1039/D0LC01012J}
  {\bibfield  {journal} {\bibinfo  {journal} {Lab Chip}\ } (\bibinfo {year}
  {2021})}\BibitemShut {NoStop}%
\bibitem [{\citenamefont {Friend}\ and\ \citenamefont
  {Yeo}(2011)}]{friend_microscale_2011}%
  \BibitemOpen
  \bibfield  {author} {\bibinfo {author} {\bibfnamefont {J.}~\bibnamefont
  {Friend}}\ and\ \bibinfo {author} {\bibfnamefont {L.~Y.}\ \bibnamefont
  {Yeo}},\ }\href {https://doi.org/10.1103/RevModPhys.83.647} {\bibfield
  {journal} {\bibinfo  {journal} {Rev. Mod. Phys.}\ }\textbf {\bibinfo {volume}
  {83}},\ \bibinfo {pages} {647} (\bibinfo {year} {2011})}\BibitemShut
  {NoStop}%
\bibitem [{\citenamefont {Plaksin}\ \emph {et~al.}(2014)\citenamefont
  {Plaksin}, \citenamefont {Shoham},\ and\ \citenamefont
  {Kimmel}}]{plaksin_intramembrane_2014}%
  \BibitemOpen
  \bibfield  {author} {\bibinfo {author} {\bibfnamefont {M.}~\bibnamefont
  {Plaksin}}, \bibinfo {author} {\bibfnamefont {S.}~\bibnamefont {Shoham}},\
  and\ \bibinfo {author} {\bibfnamefont {E.}~\bibnamefont {Kimmel}},\ }\href
  {https://doi.org/10.1103/PhysRevX.4.011004} {\bibfield  {journal} {\bibinfo
  {journal} {Phys. Rev. X}\ }\textbf {\bibinfo {volume} {4}},\ \bibinfo {pages}
  {011004} (\bibinfo {year} {2014})}\BibitemShut {NoStop}%
\bibitem [{\citenamefont {Connacher}\ \emph {et~al.}(2018)\citenamefont
  {Connacher}, \citenamefont {Zhang}, \citenamefont {Huang}, \citenamefont
  {Mei}, \citenamefont {Zhang}, \citenamefont {Gopesh},\ and\ \citenamefont
  {Friend}}]{connacher_micro/nano_2018}%
  \BibitemOpen
  \bibfield  {author} {\bibinfo {author} {\bibfnamefont {W.}~\bibnamefont
  {Connacher}}, \bibinfo {author} {\bibfnamefont {N.}~\bibnamefont {Zhang}},
  \bibinfo {author} {\bibfnamefont {A.}~\bibnamefont {Huang}}, \bibinfo
  {author} {\bibfnamefont {J.}~\bibnamefont {Mei}}, \bibinfo {author}
  {\bibfnamefont {S.}~\bibnamefont {Zhang}}, \bibinfo {author} {\bibfnamefont
  {T.}~\bibnamefont {Gopesh}},\ and\ \bibinfo {author} {\bibfnamefont
  {J.}~\bibnamefont {Friend}},\ }\href {https://doi.org/10.1039/C8LC00112J}
  {\bibfield  {journal} {\bibinfo  {journal} {Lab Chip}\ }\textbf {\bibinfo
  {volume} {18}},\ \bibinfo {pages} {1952} (\bibinfo {year}
  {2018})}\BibitemShut {NoStop}%
\bibitem [{\citenamefont {Collins}\ \emph {et~al.}(2018)\citenamefont
  {Collins}, \citenamefont {O'Rorke}, \citenamefont {Devendran}, \citenamefont
  {Ma}, \citenamefont {Han}, \citenamefont {Neild},\ and\ \citenamefont
  {Ai}}]{collins_self-aligned_2018}%
  \BibitemOpen
  \bibfield  {author} {\bibinfo {author} {\bibfnamefont {D.~J.}\ \bibnamefont
  {Collins}}, \bibinfo {author} {\bibfnamefont {R.}~\bibnamefont {O'Rorke}},
  \bibinfo {author} {\bibfnamefont {C.}~\bibnamefont {Devendran}}, \bibinfo
  {author} {\bibfnamefont {Z.}~\bibnamefont {Ma}}, \bibinfo {author}
  {\bibfnamefont {J.}~\bibnamefont {Han}}, \bibinfo {author} {\bibfnamefont
  {A.}~\bibnamefont {Neild}},\ and\ \bibinfo {author} {\bibfnamefont
  {Y.}~\bibnamefont {Ai}},\ }\href
  {https://doi.org/10.1103/PhysRevLett.120.074502} {\bibfield  {journal}
  {\bibinfo  {journal} {Phys. Rev. Lett.}\ }\textbf {\bibinfo {volume} {120}},\
  \bibinfo {pages} {074502} (\bibinfo {year} {2018})}\BibitemShut {NoStop}%
\bibitem [{\citenamefont {Connacher}\ \emph {et~al.}(2020)\citenamefont
  {Connacher}, \citenamefont {Orosco},\ and\ \citenamefont
  {Friend}}]{connacher_droplet_2020}%
  \BibitemOpen
  \bibfield  {author} {\bibinfo {author} {\bibfnamefont {W.}~\bibnamefont
  {Connacher}}, \bibinfo {author} {\bibfnamefont {J.}~\bibnamefont {Orosco}},\
  and\ \bibinfo {author} {\bibfnamefont {J.}~\bibnamefont {Friend}},\ }\href
  {https://doi.org/10.1103/PhysRevLett.125.184504} {\bibfield  {journal}
  {\bibinfo  {journal} {Phys. Rev. Lett.}\ }\textbf {\bibinfo {volume} {125}},\
  \bibinfo {pages} {184504} (\bibinfo {year} {2020})}\BibitemShut {NoStop}%
\bibitem [{\citenamefont {Lighthill}(1978)}]{lighthill_acoustic_1978}%
  \BibitemOpen
  \bibfield  {author} {\bibinfo {author} {\bibfnamefont {J.}~\bibnamefont
  {Lighthill}},\ }\href {https://doi.org/10.1016/0022-460X(78)90388-7}
  {\bibfield  {journal} {\bibinfo  {journal} {J. Sound Vib.}\ }\textbf
  {\bibinfo {volume} {61}},\ \bibinfo {pages} {391} (\bibinfo {year}
  {1978})}\BibitemShut {NoStop}%
\bibitem [{\citenamefont {Daru}\ \emph {et~al.}(2017)\citenamefont {Daru},
  \citenamefont {Reyt}, \citenamefont {Bailliet}, \citenamefont {Weisman},\
  and\ \citenamefont {{Baltean-Carl{\`e}s}}}]{daru_acoustic_2017}%
  \BibitemOpen
  \bibfield  {author} {\bibinfo {author} {\bibfnamefont {V.}~\bibnamefont
  {Daru}}, \bibinfo {author} {\bibfnamefont {I.}~\bibnamefont {Reyt}}, \bibinfo
  {author} {\bibfnamefont {H.}~\bibnamefont {Bailliet}}, \bibinfo {author}
  {\bibfnamefont {C.}~\bibnamefont {Weisman}},\ and\ \bibinfo {author}
  {\bibfnamefont {D.}~\bibnamefont {{Baltean-Carl{\`e}s}}},\ }\href
  {https://doi.org/10.1121/1.4974058} {\bibfield  {journal} {\bibinfo
  {journal} {J. Acoust. Soc. Am.}\ }\textbf {\bibinfo {volume} {141}},\
  \bibinfo {pages} {563} (\bibinfo {year} {2017})}\BibitemShut {NoStop}%
\bibitem [{\citenamefont {Bailliet}\ \emph {et~al.}(2001)\citenamefont
  {Bailliet}, \citenamefont {Gusev}, \citenamefont {Raspet},\ and\
  \citenamefont {Hiller}}]{bailliet_acoustic_2001}%
  \BibitemOpen
  \bibfield  {author} {\bibinfo {author} {\bibfnamefont {H.}~\bibnamefont
  {Bailliet}}, \bibinfo {author} {\bibfnamefont {V.}~\bibnamefont {Gusev}},
  \bibinfo {author} {\bibfnamefont {R.}~\bibnamefont {Raspet}},\ and\ \bibinfo
  {author} {\bibfnamefont {R.~A.}\ \bibnamefont {Hiller}},\ }\href
  {https://doi.org/10.1121/1.1394739} {\bibfield  {journal} {\bibinfo
  {journal} {J. Acoust. Soc. Am.}\ }\textbf {\bibinfo {volume} {110}},\
  \bibinfo {pages} {1808} (\bibinfo {year} {2001})}\BibitemShut {NoStop}%
\bibitem [{\citenamefont {Vanneste}\ and\ \citenamefont
  {B{\"u}hler}(2011)}]{vanneste_streaming_2011}%
  \BibitemOpen
  \bibfield  {author} {\bibinfo {author} {\bibfnamefont {J.}~\bibnamefont
  {Vanneste}}\ and\ \bibinfo {author} {\bibfnamefont {O.}~\bibnamefont
  {B{\"u}hler}},\ }\href {https://doi.org/10.1098/rspa.2010.0457} {\bibfield
  {journal} {\bibinfo  {journal} {P. Roy. Soc. A-Math. Phy.}\ }\textbf
  {\bibinfo {volume} {467}},\ \bibinfo {pages} {1779} (\bibinfo {year}
  {2011})}\BibitemShut {NoStop}%
\bibitem [{\citenamefont {Riaud}\ \emph {et~al.}(2017)\citenamefont {Riaud},
  \citenamefont {Baudoin}, \citenamefont {Bou~Matar}, \citenamefont {Thomas},\
  and\ \citenamefont {Brunet}}]{riaud_influence_2017}%
  \BibitemOpen
  \bibfield  {author} {\bibinfo {author} {\bibfnamefont {A.}~\bibnamefont
  {Riaud}}, \bibinfo {author} {\bibfnamefont {M.}~\bibnamefont {Baudoin}},
  \bibinfo {author} {\bibfnamefont {O.}~\bibnamefont {Bou~Matar}}, \bibinfo
  {author} {\bibfnamefont {J.-L.}\ \bibnamefont {Thomas}},\ and\ \bibinfo
  {author} {\bibfnamefont {P.}~\bibnamefont {Brunet}},\ }\href
  {https://doi.org/10.1017/jfm.2017.178} {\bibfield  {journal} {\bibinfo
  {journal} {J. Fluid Mech.}\ }\textbf {\bibinfo {volume} {821}},\ \bibinfo
  {pages} {384} (\bibinfo {year} {2017})}\BibitemShut {NoStop}%
\bibitem [{\citenamefont {Rezk}\ \emph {et~al.}(2012)\citenamefont {Rezk},
  \citenamefont {Manor}, \citenamefont {Friend},\ and\ \citenamefont
  {Yeo}}]{rezk_unique_2012}%
  \BibitemOpen
  \bibfield  {author} {\bibinfo {author} {\bibfnamefont {A.~R.}\ \bibnamefont
  {Rezk}}, \bibinfo {author} {\bibfnamefont {O.}~\bibnamefont {Manor}},
  \bibinfo {author} {\bibfnamefont {J.~R.}\ \bibnamefont {Friend}},\ and\
  \bibinfo {author} {\bibfnamefont {L.~Y.}\ \bibnamefont {Yeo}},\ }\href
  {https://doi.org/10.1038/ncomms2168} {\bibfield  {journal} {\bibinfo
  {journal} {Nat. Commun.}\ }\textbf {\bibinfo {volume} {3}},\ \bibinfo {pages}
  {1167} (\bibinfo {year} {2012})}\BibitemShut {NoStop}%
\bibitem [{\citenamefont {Tang}\ \emph {et~al.}(2017)\citenamefont {Tang},
  \citenamefont {Hu}, \citenamefont {Qian},\ and\ \citenamefont
  {Zhang}}]{tang_eckart_2017}%
  \BibitemOpen
  \bibfield  {author} {\bibinfo {author} {\bibfnamefont {Q.}~\bibnamefont
  {Tang}}, \bibinfo {author} {\bibfnamefont {J.}~\bibnamefont {Hu}}, \bibinfo
  {author} {\bibfnamefont {S.}~\bibnamefont {Qian}},\ and\ \bibinfo {author}
  {\bibfnamefont {X.}~\bibnamefont {Zhang}},\ }\href
  {https://doi.org/10.1007/s10404-017-1871-1} {\bibfield  {journal} {\bibinfo
  {journal} {Microfluid, Nanofluid.}\ }\textbf {\bibinfo {volume} {21}},\
  \bibinfo {pages} {28} (\bibinfo {year} {2017})}\BibitemShut {NoStop}%
\bibitem [{\citenamefont {Pavlic}\ and\ \citenamefont
  {Dual}(2021)}]{pavlic_streaming_2021}%
  \BibitemOpen
  \bibfield  {author} {\bibinfo {author} {\bibfnamefont {A.}~\bibnamefont
  {Pavlic}}\ and\ \bibinfo {author} {\bibfnamefont {J.}~\bibnamefont {Dual}},\
  }\bibfield  {journal} {\bibinfo  {journal} {J. Fluid Mech.}\ }\textbf
  {\bibinfo {volume} {911}},\ \href {https://doi.org/10.1017/jfm.2020.1046}
  {10.1017/jfm.2020.1046} (\bibinfo {year} {2021})\BibitemShut {NoStop}%
\bibitem [{\citenamefont {Kardar}\ \emph {et~al.}(1986)\citenamefont {Kardar},
  \citenamefont {Parisi},\ and\ \citenamefont {Zhang}}]{kardar_dynamic_1986}%
  \BibitemOpen
  \bibfield  {author} {\bibinfo {author} {\bibfnamefont {M.}~\bibnamefont
  {Kardar}}, \bibinfo {author} {\bibfnamefont {G.}~\bibnamefont {Parisi}},\
  and\ \bibinfo {author} {\bibfnamefont {Y.-C.}\ \bibnamefont {Zhang}},\ }\href
  {https://doi.org/10.1103/PhysRevLett.56.889} {\bibfield  {journal} {\bibinfo
  {journal} {Phys. Rev. Lett.}\ }\textbf {\bibinfo {volume} {56}},\ \bibinfo
  {pages} {889} (\bibinfo {year} {1986})}\BibitemShut {NoStop}%
\bibitem [{\citenamefont {Meakin}\ \emph {et~al.}(1986)\citenamefont {Meakin},
  \citenamefont {Ramanlal}, \citenamefont {Sander},\ and\ \citenamefont
  {Ball}}]{meakin_ballistic_1986}%
  \BibitemOpen
  \bibfield  {author} {\bibinfo {author} {\bibfnamefont {P.}~\bibnamefont
  {Meakin}}, \bibinfo {author} {\bibfnamefont {P.}~\bibnamefont {Ramanlal}},
  \bibinfo {author} {\bibfnamefont {L.~M.}\ \bibnamefont {Sander}},\ and\
  \bibinfo {author} {\bibfnamefont {R.~C.}\ \bibnamefont {Ball}},\ }\href
  {https://doi.org/10.1103/PhysRevA.34.5091} {\bibfield  {journal} {\bibinfo
  {journal} {Phys. Rev. A}\ }\textbf {\bibinfo {volume} {34}},\ \bibinfo
  {pages} {5091} (\bibinfo {year} {1986})}\BibitemShut {NoStop}%
\bibitem [{\citenamefont {Sasamoto}\ and\ \citenamefont
  {Spohn}(2010)}]{sasamoto_one-dimensional_2010}%
  \BibitemOpen
  \bibfield  {author} {\bibinfo {author} {\bibfnamefont {T.}~\bibnamefont
  {Sasamoto}}\ and\ \bibinfo {author} {\bibfnamefont {H.}~\bibnamefont
  {Spohn}},\ }\href {https://doi.org/10.1103/PhysRevLett.104.230602} {\bibfield
   {journal} {\bibinfo  {journal} {Phys. Rev. Lett.}\ }\textbf {\bibinfo
  {volume} {104}},\ \bibinfo {pages} {230602} (\bibinfo {year}
  {2010})}\BibitemShut {NoStop}%
\bibitem [{\citenamefont {Santalla}\ and\ \citenamefont
  {Ferreira}(2018)}]{santalla_eden_2018}%
  \BibitemOpen
  \bibfield  {author} {\bibinfo {author} {\bibfnamefont {S.~N.}\ \bibnamefont
  {Santalla}}\ and\ \bibinfo {author} {\bibfnamefont {S.~C.}\ \bibnamefont
  {Ferreira}},\ }\href {https://doi.org/10.1103/PhysRevE.98.022405} {\bibfield
  {journal} {\bibinfo  {journal} {Phys. Rev. E}\ }\textbf {\bibinfo {volume}
  {98}},\ \bibinfo {pages} {022405} (\bibinfo {year} {2018})}\BibitemShut
  {NoStop}%
\bibitem [{\citenamefont {Nahum}\ \emph {et~al.}(2017)\citenamefont {Nahum},
  \citenamefont {Ruhman}, \citenamefont {Vijay},\ and\ \citenamefont
  {Haah}}]{nahum_quantum_2017}%
  \BibitemOpen
  \bibfield  {author} {\bibinfo {author} {\bibfnamefont {A.}~\bibnamefont
  {Nahum}}, \bibinfo {author} {\bibfnamefont {J.}~\bibnamefont {Ruhman}},
  \bibinfo {author} {\bibfnamefont {S.}~\bibnamefont {Vijay}},\ and\ \bibinfo
  {author} {\bibfnamefont {J.}~\bibnamefont {Haah}},\ }\href
  {https://doi.org/10.1103/PhysRevX.7.031016} {\bibfield  {journal} {\bibinfo
  {journal} {Phys. Rev. X}\ }\textbf {\bibinfo {volume} {7}},\ \bibinfo {pages}
  {031016} (\bibinfo {year} {2017})}\BibitemShut {NoStop}%
\bibitem [{\citenamefont {Burgers}(1974)}]{burgers_nonlinear_1974}%
  \BibitemOpen
  \bibfield  {author} {\bibinfo {author} {\bibfnamefont {J.~M.}\ \bibnamefont
  {Burgers}},\ }\href {https://doi.org/10.1007/978-94-010-1745-9} {\emph
  {\bibinfo {title} {The {{Nonlinear Diffusion Equation}}: {{Asymptotic
  Solutions}} and {{Statistical Problems}}}}}\ (\bibinfo  {publisher}
  {{Springer Netherlands}},\ \bibinfo {year} {1974})\BibitemShut {NoStop}%
\bibitem [{\citenamefont {Kamakura}\ \emph {et~al.}(1996)\citenamefont
  {Kamakura}, \citenamefont {Sudo}, \citenamefont {Matsuda},\ and\
  \citenamefont {Kumamoto}}]{kamakura_time_1996}%
  \BibitemOpen
  \bibfield  {author} {\bibinfo {author} {\bibfnamefont {T.}~\bibnamefont
  {Kamakura}}, \bibinfo {author} {\bibfnamefont {T.}~\bibnamefont {Sudo}},
  \bibinfo {author} {\bibfnamefont {K.}~\bibnamefont {Matsuda}},\ and\ \bibinfo
  {author} {\bibfnamefont {Y.}~\bibnamefont {Kumamoto}},\ }\href
  {https://doi.org/10.1121/1.415948} {\bibfield  {journal} {\bibinfo  {journal}
  {J. Acoust. Soc. Am.}\ }\textbf {\bibinfo {volume} {100}},\ \bibinfo {pages}
  {132} (\bibinfo {year} {1996})}\BibitemShut {NoStop}%
\bibitem [{\citenamefont {Aln{\ae}s}\ \emph {et~al.}(2009)\citenamefont
  {Aln{\ae}s}, \citenamefont {Logg}, \citenamefont {Mardal}, \citenamefont
  {Skavhaug},\ and\ \citenamefont {Langtangen}}]{alnaes_unified_2009}%
  \BibitemOpen
  \bibfield  {author} {\bibinfo {author} {\bibfnamefont {M.~S.}\ \bibnamefont
  {Aln{\ae}s}}, \bibinfo {author} {\bibfnamefont {A.}~\bibnamefont {Logg}},
  \bibinfo {author} {\bibfnamefont {K.-A.}\ \bibnamefont {Mardal}}, \bibinfo
  {author} {\bibfnamefont {O.}~\bibnamefont {Skavhaug}},\ and\ \bibinfo
  {author} {\bibfnamefont {H.~P.}\ \bibnamefont {Langtangen}},\ }\href
  {https://doi.org/10.1504/IJCSE.2009.029160} {\bibfield  {journal} {\bibinfo
  {journal} {Int. J. Comput. Sci. Eng.}\ }\textbf {\bibinfo {volume} {4}},\
  \bibinfo {pages} {231} (\bibinfo {year} {2009})}\BibitemShut {NoStop}%
\bibitem [{\citenamefont {Aln{\ae}s}\ \emph {et~al.}(2015)\citenamefont
  {Aln{\ae}s}, \citenamefont {Blechta}, \citenamefont {Hake}, \citenamefont
  {Johansson}, \citenamefont {Kehlet}, \citenamefont {Logg}, \citenamefont
  {Richardson}, \citenamefont {Ring}, \citenamefont {Rognes},\ and\
  \citenamefont {Wells}}]{alnaes_fenics_2015}%
  \BibitemOpen
  \bibfield  {author} {\bibinfo {author} {\bibfnamefont {M.~S.}\ \bibnamefont
  {Aln{\ae}s}}, \bibinfo {author} {\bibfnamefont {J.}~\bibnamefont {Blechta}},
  \bibinfo {author} {\bibfnamefont {J.}~\bibnamefont {Hake}}, \bibinfo {author}
  {\bibfnamefont {A.}~\bibnamefont {Johansson}}, \bibinfo {author}
  {\bibfnamefont {B.}~\bibnamefont {Kehlet}}, \bibinfo {author} {\bibfnamefont
  {A.}~\bibnamefont {Logg}}, \bibinfo {author} {\bibfnamefont {C.}~\bibnamefont
  {Richardson}}, \bibinfo {author} {\bibfnamefont {J.}~\bibnamefont {Ring}},
  \bibinfo {author} {\bibfnamefont {M.~E.}\ \bibnamefont {Rognes}},\ and\
  \bibinfo {author} {\bibfnamefont {G.~N.}\ \bibnamefont {Wells}},\ }\href
  {https://doi.org/10.11588/ans.2015.100.20553} {\bibfield  {journal} {\bibinfo
   {journal} {Arch. Num. Sofwr.}\ }\textbf {\bibinfo {volume} {3}},\ \bibinfo
  {pages} {9} (\bibinfo {year} {2015})}\BibitemShut {NoStop}%
\bibitem [{\citenamefont {Aln{\ae}s}\ \emph {et~al.}(2014)\citenamefont
  {Aln{\ae}s}, \citenamefont {Logg}, \citenamefont {{\O}lgaard}, \citenamefont
  {Rognes},\ and\ \citenamefont {Wells}}]{alnaes_unified_2014}%
  \BibitemOpen
  \bibfield  {author} {\bibinfo {author} {\bibfnamefont {M.~S.}\ \bibnamefont
  {Aln{\ae}s}}, \bibinfo {author} {\bibfnamefont {A.}~\bibnamefont {Logg}},
  \bibinfo {author} {\bibfnamefont {K.~B.}\ \bibnamefont {{\O}lgaard}},
  \bibinfo {author} {\bibfnamefont {M.~E.}\ \bibnamefont {Rognes}},\ and\
  \bibinfo {author} {\bibfnamefont {G.~N.}\ \bibnamefont {Wells}},\ }\href
  {https://doi.org/10.1145/2566630} {\bibfield  {journal} {\bibinfo  {journal}
  {ACM Trans. Math. Softw.}\ }\textbf {\bibinfo {volume} {40}},\ \bibinfo
  {pages} {9:1} (\bibinfo {year} {2014})}\BibitemShut {NoStop}%
\bibitem [{\citenamefont {Kirby}(2004)}]{kirby_algorithm_2004}%
  \BibitemOpen
  \bibfield  {author} {\bibinfo {author} {\bibfnamefont {R.~C.}\ \bibnamefont
  {Kirby}},\ }\href {https://doi.org/10.1145/1039813.1039820} {\bibfield
  {journal} {\bibinfo  {journal} {ACM Trans. Math. Softw.}\ }\textbf {\bibinfo
  {volume} {30}},\ \bibinfo {pages} {502} (\bibinfo {year} {2004})}\BibitemShut
  {NoStop}%
\bibitem [{\citenamefont {Kirby}\ and\ \citenamefont
  {Logg}(2006)}]{kirby_compiler_2006}%
  \BibitemOpen
  \bibfield  {author} {\bibinfo {author} {\bibfnamefont {R.~C.}\ \bibnamefont
  {Kirby}}\ and\ \bibinfo {author} {\bibfnamefont {A.}~\bibnamefont {Logg}},\
  }\href {https://doi.org/10.1145/1163641.1163644} {\bibfield  {journal}
  {\bibinfo  {journal} {ACM Trans. Math. Softw.}\ }\textbf {\bibinfo {volume}
  {32}},\ \bibinfo {pages} {417} (\bibinfo {year} {2006})}\BibitemShut
  {NoStop}%
\bibitem [{\citenamefont {Logg}\ \emph {et~al.}(2012)\citenamefont {Logg},
  \citenamefont {Mardal},\ and\ \citenamefont {Wells}}]{logg_automated_2012}%
  \BibitemOpen
  \bibinfo {editor} {\bibfnamefont {A.}~\bibnamefont {Logg}}, \bibinfo {editor}
  {\bibfnamefont {K.-A.}\ \bibnamefont {Mardal}},\ and\ \bibinfo {editor}
  {\bibfnamefont {G.~N.}\ \bibnamefont {Wells}},\ eds.,\ \href
  {https://doi.org/10.1007/978-3-642-23099-8} {\emph {\bibinfo {title}
  {Automated {{Solution}} of {{Differential Equations}} by the {{Finite Element
  Method}}: {{The FEniCS Book}}}}},\ Lecture {{Notes}} in {{Computational
  Science}} and {{Engineering}}\ (\bibinfo  {publisher} {{Springer-Verlag}},\
  \bibinfo {address} {{Berlin Heidelberg}},\ \bibinfo {year}
  {2012})\BibitemShut {NoStop}%
\bibitem [{\citenamefont {Logg}\ and\ \citenamefont
  {Wells}(2010)}]{logg_dolfin_2010}%
  \BibitemOpen
  \bibfield  {author} {\bibinfo {author} {\bibfnamefont {A.}~\bibnamefont
  {Logg}}\ and\ \bibinfo {author} {\bibfnamefont {G.~N.}\ \bibnamefont
  {Wells}},\ }\href {https://doi.org/10.1145/1731022.1731030} {\bibfield
  {journal} {\bibinfo  {journal} {ACM Trans. Math. Softw.}\ }\textbf {\bibinfo
  {volume} {37}},\ \bibinfo {pages} {20:1} (\bibinfo {year}
  {2010})}\BibitemShut {NoStop}%
\bibitem [{\citenamefont {{\O}lgaard}\ and\ \citenamefont
  {Wells}(2010)}]{olgaard_optimizations_2010}%
  \BibitemOpen
  \bibfield  {author} {\bibinfo {author} {\bibfnamefont {K.~B.}\ \bibnamefont
  {{\O}lgaard}}\ and\ \bibinfo {author} {\bibfnamefont {G.~N.}\ \bibnamefont
  {Wells}},\ }\href {https://doi.org/10.1145/1644001.1644009} {\bibfield
  {journal} {\bibinfo  {journal} {ACM Trans. Math. Softw.}\ }\textbf {\bibinfo
  {volume} {37}},\ \bibinfo {pages} {8:1} (\bibinfo {year} {2010})}\BibitemShut
  {NoStop}%
\bibitem [{\citenamefont {Rudenko}\ and\ \citenamefont
  {Soluyan}(1977)}]{rudenko_theoretical_1977}%
  \BibitemOpen
  \bibfield  {author} {\bibinfo {author} {\bibfnamefont {O.~V.}\ \bibnamefont
  {Rudenko}}\ and\ \bibinfo {author} {\bibfnamefont {S.~I.}\ \bibnamefont
  {Soluyan}},\ }\href@noop {} {\emph {\bibinfo {title} {Theoretical Foundations
  of Nonlinear Acoustics}}},\ Studies in Soviet Science\ (\bibinfo  {publisher}
  {{Plenum}},\ \bibinfo {address} {{New York}},\ \bibinfo {year}
  {1977})\BibitemShut {NoStop}%
\bibitem [{\citenamefont {Zabolotskaya}\ and\ \citenamefont
  {Khokhlov}(1969)}]{zabolotskaya_quasiplane_1969}%
  \BibitemOpen
  \bibfield  {author} {\bibinfo {author} {\bibfnamefont {E.~A.}\ \bibnamefont
  {Zabolotskaya}}\ and\ \bibinfo {author} {\bibfnamefont {R.~V.}\ \bibnamefont
  {Khokhlov}},\ }\href@noop {} {\bibfield  {journal} {\bibinfo  {journal} {Sov.
  Phys. Acoust.}\ }\textbf {\bibinfo {volume} {15}},\ \bibinfo {pages} {35}
  (\bibinfo {year} {1969})}\BibitemShut {NoStop}%
\bibitem [{\citenamefont {Kuznetsov}(1971)}]{kuznetsov_equations_1971}%
  \BibitemOpen
  \bibfield  {author} {\bibinfo {author} {\bibfnamefont {V.~P.}\ \bibnamefont
  {Kuznetsov}},\ }\href@noop {} {\bibfield  {journal} {\bibinfo  {journal}
  {Sov. Phys. Acoust.}\ }\textbf {\bibinfo {volume} {16}},\ \bibinfo {pages}
  {467} (\bibinfo {year} {1971})}\BibitemShut {NoStop}%
\bibitem [{\citenamefont {Novikov}\ \emph {et~al.}(1987)\citenamefont
  {Novikov}, \citenamefont {Rudenko},\ and\ \citenamefont
  {Timoshenko}}]{novikov_nonlinear_1987}%
  \BibitemOpen
  \bibfield  {author} {\bibinfo {author} {\bibfnamefont {B.~K.}\ \bibnamefont
  {Novikov}}, \bibinfo {author} {\bibfnamefont {O.~V.}\ \bibnamefont
  {Rudenko}},\ and\ \bibinfo {author} {\bibfnamefont {V.~I.}\ \bibnamefont
  {Timoshenko}},\ }\href@noop {} {\emph {\bibinfo {title} {Nonlinear
  {{Underwater Acoustics}}}}}\ (\bibinfo  {publisher} {{American Institute of
  Physics}},\ \bibinfo {year} {1987})\BibitemShut {NoStop}%
\bibitem [{\citenamefont {McGough}\ \emph {et~al.}(2004)\citenamefont
  {McGough}, \citenamefont {Samulski},\ and\ \citenamefont
  {Kelly}}]{mcgough_efficient_2004}%
  \BibitemOpen
  \bibfield  {author} {\bibinfo {author} {\bibfnamefont {R.~J.}\ \bibnamefont
  {McGough}}, \bibinfo {author} {\bibfnamefont {T.~V.}\ \bibnamefont
  {Samulski}},\ and\ \bibinfo {author} {\bibfnamefont {J.~F.}\ \bibnamefont
  {Kelly}},\ }\href {https://doi.org/10.1121/1.1687835} {\bibfield  {journal}
  {\bibinfo  {journal} {The Journal of the Acoustical Society of America}\
  }\textbf {\bibinfo {volume} {115}},\ \bibinfo {pages} {1942} (\bibinfo {year}
  {2004})}\BibitemShut {NoStop}%
\bibitem [{\citenamefont {Chen}\ and\ \citenamefont
  {McGough}(2008)}]{chen_2d_2008}%
  \BibitemOpen
  \bibfield  {author} {\bibinfo {author} {\bibfnamefont {D.}~\bibnamefont
  {Chen}}\ and\ \bibinfo {author} {\bibfnamefont {R.~J.}\ \bibnamefont
  {McGough}},\ }\href {https://doi.org/10.1121/1.2950081} {\bibfield  {journal}
  {\bibinfo  {journal} {J. Acoust. Soc. Am.}\ }\textbf {\bibinfo {volume}
  {124}},\ \bibinfo {pages} {1526} (\bibinfo {year} {2008})}\BibitemShut
  {NoStop}%
\bibitem [{\citenamefont {McGough}(2021)}]{mcgough_fast_2021}%
  \BibitemOpen
  \bibfield  {author} {\bibinfo {author} {\bibfnamefont {R.~J.}\ \bibnamefont
  {McGough}},\ }\href {https://www.egr.msu.edu/~fultras-web/index.php}
  {\bibinfo {title} {Fast {{Object}}-{{Oriented C}}++ {{Ultrasound Simulator}}
  ({{FOCUS}})}},\ \bibinfo {howpublished} {Michigan State University} (\bibinfo
  {year} {2021})\BibitemShut {NoStop}%
\bibitem [{\citenamefont {Riccati}(1724)}]{riccati_animadversiones_1724}%
  \BibitemOpen
  \bibfield  {author} {\bibinfo {author} {\bibfnamefont {J.}~\bibnamefont
  {Riccati}},\ }\href@noop {} {\bibfield  {journal} {\bibinfo  {journal} {Acta
  Erud.}\ }\textbf {\bibinfo {volume} {8}},\ \bibinfo {pages} {66} (\bibinfo
  {year} {1724})}\BibitemShut {NoStop}%
\bibitem [{\citenamefont {Moudjed}\ \emph {et~al.}(2015)\citenamefont
  {Moudjed}, \citenamefont {Botton}, \citenamefont {Henry}, \citenamefont
  {Millet}, \citenamefont {Garandet},\ and\ \citenamefont
  {Ben~Hadid}}]{moudjed_near-field_2015}%
  \BibitemOpen
  \bibfield  {author} {\bibinfo {author} {\bibfnamefont {B.}~\bibnamefont
  {Moudjed}}, \bibinfo {author} {\bibfnamefont {V.}~\bibnamefont {Botton}},
  \bibinfo {author} {\bibfnamefont {D.}~\bibnamefont {Henry}}, \bibinfo
  {author} {\bibfnamefont {S.}~\bibnamefont {Millet}}, \bibinfo {author}
  {\bibfnamefont {J.~P.}\ \bibnamefont {Garandet}},\ and\ \bibinfo {author}
  {\bibfnamefont {H.}~\bibnamefont {Ben~Hadid}},\ }\href
  {https://doi.org/10.1103/PhysRevE.91.033011} {\bibfield  {journal} {\bibinfo
  {journal} {Phys. Rev. E}\ }\textbf {\bibinfo {volume} {91}},\ \bibinfo
  {pages} {033011} (\bibinfo {year} {2015})}\BibitemShut {NoStop}%
\bibitem [{\citenamefont {Ince}(1956)}]{ince_ordinary_1956}%
  \BibitemOpen
  \bibfield  {author} {\bibinfo {author} {\bibfnamefont {E.}~\bibnamefont
  {Ince}},\ }\href@noop {} {\emph {\bibinfo {title} {Ordinary {{Differential
  Equations}}}}}\ (\bibinfo  {publisher} {{Dover}},\ \bibinfo {address} {{New
  York}},\ \bibinfo {year} {1956})\BibitemShut {NoStop}%
\bibitem [{\citenamefont
  {Holoborodko}(2020)}]{holoborodko_multiprecision_2020}%
  \BibitemOpen
  \bibfield  {author} {\bibinfo {author} {\bibfnamefont {P.}~\bibnamefont
  {Holoborodko}},\ }\href {https://www.advanpix.com/} {\bibinfo {title}
  {Multiprecision {{Computing Toolbox}} for {{MATLAB}}}},\ \bibinfo
  {howpublished} {Advanpix LLC} (\bibinfo {year} {2020})\BibitemShut {NoStop}%
\end{thebibliography}%

\end{document}